\documentstyle[aps,preprint]{revtex}
\input epsf

\begin{document}

\preprint{\vbox{\hspace*{\fill} DOE/ER/40762-115\\
          \hspace*{\fill} U. of MD PP \#97-087}} \vspace{.5in}

\title{Comment on "A Search for Narrow Sum Energy Lines in Electron-Positron Pair Emission from Heavy Ion collisions near the Coulomb Barrier" }

\author{James J. Griffin}

\address{Department of 
Physics, University of~Maryland, College~Park, MD~20742}

\maketitle

 
\vspace*{\fill}
\noindent{submitted to Phys. Rev. Letters; posted as nucl-th/9703041 at xxx.lanl.gov}
\noindent{Typeset using REVTEX}

\newpage
\narrowtext
The report of Ahmad et al.\cite{ahma/95}, states that ``No evidence is
found for sharp peaks in the present data.'' But this statement is
factually inaccurate. Figure 1 exhibits\footnote{We are grateful to the
APEX collaboration for supplying in numerical form the data graphed in
Figure 2 of Ref.\cite{ahma/95}.} the APEX' published U + Th ``wedge
cut'' data, and background, from  Ref.\cite{ahma/95}. These data show a
distinct maximum of 83 counts (3.2$\sigma$) in the pair excess near 790
keV. Our one-gaussian best fit\cite{grif/97u2} to this excess reduces
$\chi^2$ (for the 21 bins from 500 to 920 keV) by 12.6 from 28.3 to
16.7, and $\chi^2$/D from 1.41 to 0.98.  Furthermore, at the 99\% CL,
the data imposes lower and upper bounds, ($\nu_L,\nu_U$) = (24, 146)
upon the mean excess pair count in this bin. This data provides
positive, statistically significant evidence for sharp pairs near 790
keV. (Whether this evidence is compelling is a separate question  not
addressed here.)

Ref\cite{ahma/95} also presents a misleading view of the EPOS
data\cite{cowa/86,sala/90} by epitomizing those results by a constant
5.0$\mu$b/sr sharp pair cross section, and by inferring therefrom that
the APEX experiment should record about 2500 sharp pairs near 800 keV
(Fig. 2a of Ref.\cite{ahma/95}). But Table I shows that if the cross
section were 5.0$\mu$b/sr and constant in beam energy as assumed, then
EPOS would have counted more than 900 sharp pairs, and not the
$\sim$100 they reported.  Thus, the APEX expectations are exaggerated
by $\sim$10$\times$; APEX should have expected not $\sim$2500, but $\sim$250
sharp pairs.

This exaggeration results because the EPOS' published 5.0$\mu$b/sr
cross section is not a correct average over EPOS' 0.07 MeV/U beam
spread.  In fact, EPOS' $\sim$100 sharp pair count defines an {\it
energy integrated cross section} of about 0.09($\mu$b/sr)(MeV/U), so
that their correct\footnote{Ganz et al.\cite{ganz/96} report a revised
value:$\sim$1.4$\mu$b/sr.} average cross section is $\sim$1.3$\mu$b/sr, 
not 5.0$\mu$b/sr.  In

\narrowtext
\begin{flushleft}
\renewcommand{\arraystretch}{1.0}
\begin{tabular}{|l|lc|c|}  \hline 
\multicolumn{4}{|c|}{TABLE I: EXPECTED SHARP PAIR COUNTS}\\ \hline \hline
\multicolumn{4}{|l|}{CONSTANT AVERAGE CROSS SECTION:}\\ 
\multicolumn{1}{|l}{}&\multicolumn{2}{c}{$\overline{<d\sigma/d\Omega>}$ ($L^{TOT}\Delta\Omega_{2I}^{eff}G_{SP})$}&\multicolumn{1}{c|}{=N$^{EXP}_{SP}$}\\ \hline
1.APEX:    &5.0$\mu$b/sr   &492.8sr/$\mu$b  &2464 \\ \hline
2.EPOS:    &5.0   &187.8  &939  \\ \hline
\multicolumn{4}{|l|}{($L^{TOT}\Delta\Omega_{2I}^{eff}G_{SP}$)=(7000)(6.86)(0.0103)=492.8 (APEX)}\\
\multicolumn{4}{|l|}{\hspace*{30mm}=(2196)(7.03)(0.0122)=187.8 (EPOS)}\\ 
\multicolumn{4}{|l|}{\hspace*{5mm} and where G$_{SP}$ = ($\epsilon_{SP}$)(LT)(W$_{SP}$)=0.0103(APEX)}\\
\multicolumn{4}{|l|}{\hspace* {55mm} =0.0122(EPOS)}\\ \hline
\end{tabular}

\narrowtext
\footnotesize
Table I displays the numbers of sharp pair counts expected in the the
APEX and the EPOS experiments under  APEX' constant 5.0$\mu$b/sr
assumption.  The Table shows that the assumption is inconsistent with
the EPOS' $\sim$100 sharp pairs by about 10$\times$. The Table also summarizes
the parameters of these calculations, (Luminosity, Effective ion solid
angle, and Gathering Power = (Detection efficiency)x(LiveTime)x(Wedge
Factor)), and indicates their values. The units are $\mu$b, and sr. See
Ref\cite{grif/97u2} for a more detailed discussion.
\end{flushleft}
\renewcommand{\baselinestretch}{1} 
\normalsize


\narrowtext


\narrowtext
\noindent   addition APEX' average value is smaller,  by the beam
spread ratio, (0.07/0.17), than EPOS'.  Then the correct expected
average sharp pair cross section APEX' case is
$\sim$0.5$\mu$b/sr, about 10$\times$ smaller than APEX assumed.

This value is smaller than APEX' upper bound\cite{ahma/95}
(0.57$\mu$b/sr at 800 keV). Thus the APEX data cannot exclude the
pair signal which ought to have been expected. Also, the best fit of Fig.1
to APEX' data yields 123$\pm$46 pairs, a value within APEX'
empirical upper bound.

Finally we consider a strictly empirical comparison of the APEX and
EPOS pair data sets. APEX actually counted\cite{wola/95}\ a total of
40.8K (and a density of 1480 per 20 keV near 800 keV) RL(1,1) pairs of
the type which EPOS accepted: pairs with one lepton in each hemisphere
and which originate in a (1,1) event, in which one positron and one and
only one electron, are recorded. EPOS counted 50K (and 1280 per 20 keV
near 800 keV) such pairs. Thus from a purely empirical viewpoint, APEX'
expected $\sim$2500 sharp pairs in its 40.8K backround (6\%), is
grossly disproportionate to EPOS' $\sim$100 sharp pairs in its 50K
backround (0.2\%).

In summary, a)the outlook of the APEX report is distorted by its
excessive expectations, and b) the report overlooks its own positive
evidence for 123 sharp pairs.

\subsection{FIGURE CAPTION}
\noindent FIG.1. The APEX data exhibit a peak near 790 keV.


\end{document}